\documentclass[twocolumn,showpacs,preprintnumbers,amsmath,amssymb, prl]{revtex4}
\usepackage{graphicx}

\begin{document}

\title {Is the mean-field approximation so bad?\\
A simple generalization yelding realistic critical indices for 3D
Ising-class systems}

\author{A.N. Rubtsov}

\email{alex@shg.ru}

\affiliation {Physical Department of Moscow State University,
 Moscow 119899, Russia}

\date{\today}

\begin{abstract}
Modification of the renormalization-group approach, invoking
Stratonovich transformation at each step, is proposed to describe
phase transitions in 3D Ising-class systems. The proposed method
is closely related to the mean-field approximation. The low-order
scheme works well for a wide thermal range, is consistent with a
scaling hypothesis and predicts very reasonable values of
critical indices.
\end{abstract}

\pacs{05.70.Fh}

\maketitle

It is well understood since 1960's that an essential increase of
fluctuations accompanies a second-order phase transition. Exact
results for 2D Ising model \cite{Ising2D} showed that Landau
phenomenology \cite{Landau} was too crude to describe the vicinity
of the transition point. Ginzburg demonstrated that it is
correlations what breaks Landau approach down \cite{Ginzburg}. He
established the parameter which determines the thermal range for
the so-called fluctuation region.
According to the scaling hypothesis \cite{PP}, the properties of
a system in the fluctuation region are determined by the single
quantity $r_c$, being the correlation length for fluctuations.
This parameter shows a power-law dependence on the external field
$h$ and on the thermal interval  $t$ from the transition:
$r_c\propto h^{-\mu}$, $r_c\propto t^{-\nu}$. Values of $\mu$ and
$\nu$ are determined by the universal characteristics, like the
dimensionality $D$, the number of components of the order
parameter, and the type of the interaction in the system
(long-range or short-range). Other critical indices can be
expressed via $\mu$ and $\nu$. For instance for the scalar order
parameter $\bar{x}$ one obtains $\bar{x}\propto t^{-\beta},
\beta=\nu D - \nu/\mu$.

The general framework to obtain critical indices is the
renormalization-group (RG) approach \cite{Wilson,RevRG,ERGE}. The
RG transformation consists of the two stages. First, one
integrates out all $x_k$ with wave vectors $k>k_0/s$, where $k_0$
is a cut-off and $s>1$ is a parameter. The main assumption of the
method is that the potential energy of thus renormalized system
can be truncated to certain simplified form, for example to the
form of $\phi^4$ model. The second step is a scale transformation.
It is aimed to restore the cut-off for $k$ back to $k_0$ and
preserve the original dispersion law for the long-wavelength
modes.

The RG transformation should be applied recursively; its unstable
fixed point corresponds to a phase transition.  Lyapunov exponents
for the vicinity of the fixed point determine the critical
indices. The scaling hypothesis requires exactly two of the
Lyapunov exponents to be positive, they correspond to the two
independent critical indices.

With respect to the value of $s$, two modifications of the RG
scheme can be distinguished. Wilson's $\varepsilon$ expansion
\cite{Wilson, RevRG} deals with $s \gg 1$. The method is formally
valid near the upper critical dimension $D_{up}$ (for the
Ising-class classical systems $D_{up}=4$).The ''most divergent``
diagram series can be pointed out in this case. This gives the
asymptotic expansion in terms of $\varepsilon\equiv D_{up}-D$. In
the continuous version of the scheme (so-called exact RG approach)
\cite{ERGE}, $s$ is infinitesimally close to unity. Certain
decoupling (truncation) should be introduced in this case to solve
the integro-differential equation for RG flow. An accuracy is then
determined by a form of the decoupling used.

All the approaches mentioned above are very well-developed. They
allow to calculate critical indices and universal combinations of
amplitudes with a high accuracy. Progress is also achieved in the
calculation of non-universal quantities; particularly a crossover
between Gaussian (Landau-like) and critical behaviour is analyzed
\cite{crossover}. However RG is applicable only for
long-wavelength excitations since they can be described as a
fluctuating field. Therefore, in a practical calculation,
short-wavelength excitations should be first integrated out by
means of some perturbation expansion, and then RG should be
applied. The short- and the long-wavelength excitations should be
separated {\it ad hoc}. Such a two-step procedure is not very
elegant. At least from the methodological point of view it is
desired to develop a scheme which gives all the information in a
uniform manner.

It is interesting to note that calculations are much simpler
outside the critical region. In fact, even the mean-field
approximation (MFA) is usually sufficient for the qualitative
analysis of this region. In MFA, the non-local interaction in the
system is replaced by an effect of a self-consistent average
field. The MFA predictions for the transition temperature and
thermal behaviour of the order parameter in 3D case are quite
reasonable \cite{MFA}. On the other hand MFA gives Landau set of
indices and is not very accurate. Note that Landau approach itself
has a mean-field nature.

In this paper, we present a simple generalization of MFA,
performing well for the 3D Ising-class systems both nearby and far
from the transition. Particularly the critical behaviour is
reproduced correctly: the values of indices are predicted with a
few-percent accuracy. Essentially, the present approach consists
in the alternating application of the Stratonovich transformation
and the most primitive version of the renormalization-group
transformation at a finite value of $s$.

We consider a classical scalar field with an anharmonic local
potential and a short-range harmonic interaction at temperature
$T$. The partition function of this system is
\begin{eqnarray}\label{Z0}
  Z&=&\int [Dx] e^{-V/T}; \\  \nonumber
  V&=&\int U(x_r+\bar{x}) d {\bf r}+\sum_k \frac{\Omega_k
  |x_k|^2}{2}.
\end{eqnarray}
Here $U(x+\bar{x})$ is an even on-site potential, $x_k$ is a
Fourier transform of the displacement $x_r$ from the average
position $\bar{x}$; $\bar{x} \neq 0$ below the transition
temperature, and the average of $x_r$ equals zero by definition.
To make the definitions unambiguous, we put additional conditions
$U(\bar{x})=0;\Omega_{k=0}=0$. The wave vector $k$ does not exceed
the unititary cut-off: $k^2 \le 1$. The case of a short-range
interaction corresponds to the asymptotic behaviour
\begin{equation} \label{om}
\Omega_k = \omega k^2, k \to 0.
\end{equation}

The simplest way to integrate out the modes is to skip $x_k$ with
$k>1/s$. This corresponds to the zero-th approximation in the
nonlinearity of $U$. The subsequent scaling results in the RG
transformation $k \to s k, U \to s^3 U, x \to x/\sqrt{s}$. The
defect of thus applied RG method comes from the the increase of
correlations (non-linear part of $U$ grows). Therefore, after
several iterative transformations, the RG procedure is not valid
any more. Although the formal analysis of the fixed point $U(x)=0$
can be done, it wrongly gives Landau-like critical indices.

To avoid the problem higher-order approximations and the
truncation of the potential are used in known schemes described
above (a very clear explanation is given in \cite{fermi}). Here we
use another procedure: the zero-th approximation is used, but the
change of variables is made at each RG step. The change of
variables reduces correlations and is aimed to compensate its
increase at RG transformation.

At the first stage, we integrate out (omit) all the modes with
$\Omega_k>\bar{\Omega}$, where $\bar{\Omega}=\frac{3}{4 \pi} \int
\Omega_k d^3 k$ is an average of $\Omega_k$.  For the rest of the
modes, the Stratonovich identity is utilized:
\begin{widetext}
\begin{equation}\label{Str}
\exp\left(-\frac{(\Omega_k-\bar{\Omega})
|x_k|^2}{T}\right)=\frac{\bar{\Omega}-\Omega_k}{\pi
\bar{\Omega}^2} \int \exp \left(  -\frac{1}{T}
  \left( 2 \bar{\Omega} {\rm Re}(x_k
f_k^*)+\frac{\bar{\Omega} \Omega_k |f_k|^2}{\bar{\Omega}-\Omega_k}
\right) \right)df_k.
\end{equation}
Here the integration over the complex variable $f_k=\Re(f_k)+i
\Im(f_k) $ denotes the integration over the complex plane: $\int
df_k \equiv \int_{-\infty}^{\infty} d \Re(f_k)
\int_{-\infty}^{\infty} d \Im(f_k)$.

After (\ref{Str}) is substituted into (\ref{Z0}), expressions
$\sum_k {\rm Re}(x_k f_k^*) = \int x_r f_r d \bf {r}$ and $\sum_k
\bar{\Omega} |x_k|^2 = \int \bar{\Omega} x_r^2$ can be collected
in the exponent. Therefore
\begin{equation}
  Z\propto \int [Dx] \int [Df] \exp\left(-\frac{1}{T}
  \left(\int \left(U(x_r+\bar{x})-\Omega_k f_r x_r + \frac{\bar{\Omega} x_r^2}{2}\right) d {\bf r}
  +\sum_k \frac{\bar{\Omega} \Omega_k |f_k|^2}{
  2 (\bar{\Omega}-\Omega_k)}\right)\right),
\end{equation}
where $f_r$ is the inverse Fourier transform of $f_k$.

Now, we can integrate over $x_r$. Let us introduce the function
$F(f)$ accordingly to the equation:
\begin{equation}\label{Fdef}
  \exp \left(-\frac{F(f+\bar{x})-F_0}{T}\right)= \int \exp
  \left(-\frac{1}{T}\left(U(x+\bar{x})+\frac{\bar{\Omega}(x-f)^2}{2}\right)\right) dx,
\end{equation}
\end{widetext}
where $F_0$ is a constant, defined from the condition
$F(\bar{x})=0$.

With thus defined $F$, the partition function takes the form
\begin{eqnarray}\label{ZF}
    Z&\propto & \int [Df] e^{-W/T}; \\  \nonumber
    W&=&\int F(f_r+\bar{x}) d{\bf r}
  +\sum_k \frac{\bar{\Omega} \Omega_k |f_k|^2}{
  2 (\bar{\Omega}-\Omega_k)}.
\end{eqnarray}

Equation (\ref{ZF}) is formally very similar to Eq.(\ref{Z0}).
Function $F$ is even; the dispersion at $k \to 0$ coincides that
of Eq.(\ref{Z0}). One can also guess from Eq.(\ref{Str}) that as
the average of $x_r$ equals zero, the average of $f_r$ also
vanishes. Indeed, if this average would not be zero, $f_{k=0}$
would fluctuate around a macroscopically large average value, and
(\ref{Str}) would not be fulfilled.

We argue here that variables $f_k$ are much less correlated, than
$x_k$. Let us consider the parabolic approximation for $F$:
$F(\bar{x}+f)\approx \frac{\partial F(\bar{x})}{\partial f}  f +
\frac{\partial^2 F(\bar{x})}{2 \partial f^2}  f^2$. Variables
$f_k$ are uncorrelated in this approximation. Obviously $<f>$
vanishes only if $\frac{\partial F(\bar{x})}{\partial f}=0$, that
gives
\begin{equation}\label{MF}
  \frac{\bar{\Omega}\bar{x}}{T}=
  \frac{\int x \exp  \left(-T^{-1}\left(U(x+\bar{x})+\frac{1}{2}\bar{\Omega}x^2 \right)\right) dx}
  {\int \exp  \left(-T^{-1}\left(U(x+\bar{x})+\frac{1}{2}\bar{\Omega}x^2 \right)\right)
  dx}.
\end{equation}

This is exactly the mean-field equation of state for the system
(\ref{Z0}). So, even the complete neglect of correlations in $f_k$
still gives something reasonable. On the other hand, the parabolic
approximation for $U$ directly in Eq.(\ref{Z0}) gives non-sense:
since $U(x)$ does not depend on temperature, the system does not
show a transition at all in this approximation. Such a comparison
convinces us that correlations in $f_k$ are less essential, than
in $x_k$. As it was pointed above we expect that this can
compensate an increase of correlations in RG transformation.

Now, we finish the RG transformation, operating with
Eq.(\ref{ZF}). All $f_k$ with $k>1/s$ are to be integrated out
(omitted in our approximation). After the scale transformation $k
\to s k, f \to x/\sqrt{s}$ is done, the potential energy of the
system takes the form
\begin{equation}
  \int U^{(1)}(x_r+\bar{x}^{(1)}) d {\bf r}+\sum_k \frac{\Omega^{(1)}_k
  |x_k|^2}{2},
\end{equation}
where
\begin{equation} \label{Trans_k}
    \Omega^{(1)}_k=\frac{s^2 \bar{\Omega}
    \Omega_k}{\bar{\Omega}-\Omega_{k/s}},
\end{equation}

\begin{equation} \label{Trans_w}
    U^{(1)}(x+\bar{x}^{(1)})=s^3 F(x/\sqrt{s}+\bar{x}), ~~\bar{x}^{(1)}=\bar{x} \sqrt{s}.
\end{equation}

Such a transformation should be applied recursively, like in the
standard RG approach. The analysis can be easily performed
numerically.

The sequence $\Omega_k, \Omega_k^{(1)}, \Omega_k^{(2)}, ...$ is
independent of $U$ and $T$. At every step a new
$\bar{\Omega}^{(n)}$ are to be calculated. After several
transformations $\bar{\Omega}^{(n)}$ and $\Omega_k^{(n)}$ converge
to certain limit. Calculation from formula (\ref{Trans_k}) gives
\begin{equation}
  \Omega_k^{(\infty)}=\frac{\omega \bar{\Omega}^{(\infty)} (s^2-1) k^2}
  {\bar{\Omega}^{(\infty)} (s^2-1)-\omega k^2};
\end{equation}
$\bar{\Omega}^{(\infty)}$ is implicitly defined from the equation
$\bar{\Omega}^{(\infty)}=\frac{3}{4 \pi} \int \Omega_k^{(\infty)}
d^3 k$.


Now, let us consider the properties of the sequence $U, U^{(1)},
U^{(2)},...$ at $\bar{x}=0$. Transformation $U^{(n)} \to
U^{(n+1)}$ has a non-trivial unstable fixed point $U_f(x)$. This
function is shown in the upper panel of Fig.1. Consider small
deviations from this fixed point: $U^{(n)}(x)=U_f(x)+u_n(x)$. The
linearization of the transformation (\ref{Trans_w}) gives
\begin{equation}\label{lin}
  u_{n+1}(y)=u^0+s^3   \int u_n(x) A(x,y) dx;
\end{equation}
$$A=\exp\left(- \frac{U_f(x)- s^{-3} U_f(y) + F + \frac{\bar{\Omega}}{2} (x-y/\sqrt{s})^2}{T}\right).$$
where $u^0$ stands to fulfill the condition $u_{n+1}(0)=0$. The
analysis of this operator shows that among its eigenvalues only
two exceed unity. Denote them $s^{\lambda_e}$ and $s^{\lambda_o}$;
they correspond to an even and odd eigenvector, respectively. Thus
the theory is consistent with the scaling hypothesis, as it was
mentioned in an introductory part.

\begin{figure}
\includegraphics{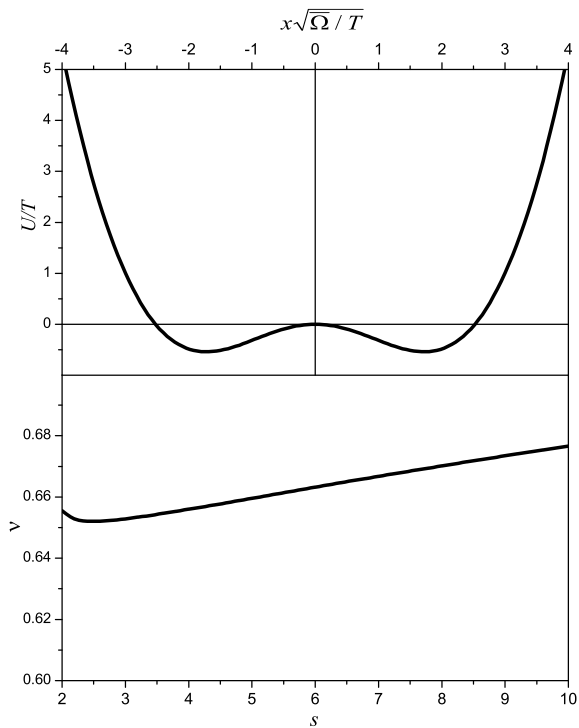}
\caption{\label{fig:epsart} Upper panel: the fixed point of the
transformation (\ref{Trans_w}) at $s=2.5$. Lower panel: critical
index $\nu$ {\it vs.} scaling factor $s$. Note a very weak
dependence of $\nu(s)$.}
\end{figure}

Exponents $\lambda_e$ and $\lambda_o$ are the inverse of the
critical indices $\nu$ and $\mu$, respectively \cite{RevRG,
fermi}. In the ideal case, critical indices should be independent
of the particular $U(x)$ and $\Omega_k$, as well as of the value
of the factor $s$. Indeed, after a proper re-scaling of the units,
the values of $T$ and $\bar{\Omega}$ drop out from
Eq.(\ref{Trans_w}). Therefore in our model $\nu$ and $\mu$ can
depend only on $s$. Further, it turns out that $\mu=2/5$ at any
$s$, as it should be for the local-potential approximation
\cite{ERGE}. What is about $\nu$, it depends on $s$ very weakly.
The numerical result for $\nu(s)$ is shown in the lower panel of
Fig.1.

On the other hand it is important to point out that the crude
universal properties of the model enter the transformation
(\ref{Trans_w}) essentially. The behaviour $\Omega_{k\to 0}
\propto k^2$ corresponds to a short-range range interaction. Such
behaviour and the dimensionality of the model determine the scale
transformation, {\it i.e.} powers of $s$ in Eq.(\ref{Trans_w}). It
is also directly reflected in the formula that the order parameter
is a scalar.

For the more detailed analysis it is reasonable to consider an
extreme value of $\nu(s)$, because in this case small variations
of $s$ do not affect $\nu$. This gives $s \approx 2.5, ~~\nu
\approx 0.652$. The obtained values of indices are collected in
Table 1. An agreement with the numerical results for the 3D Ising
model is even better than one could expect. Since the scaling
hypothesis holds in the model, the values of all other critical
indices can be determined and should have a small error bar. As an
example, the calculated value of $\beta$ is presented in Table~1.

\begin{table}
\caption{
Values of the critical indices obtained from the Lyapunov
exponents for the transformation (\ref{lin}) at $s=2.5$. Results
of the Landau theory and known values \cite{RevRG} for 3D Ising
model are given for the comparison.}
\begin{ruledtabular}
\begin{tabular}{lccc}
Index& Estimation from Eq.(\ref{lin}) &  3D Ising & Landau theory\\
$\nu$& 0.652  & 0.63 & 1/2\\
$\mu$ & 2/5 & 0.403 & 1/3\\
$\beta$& 0.326 & 0.327 & 1/2\\
\end{tabular}
\end{ruledtabular}
\end{table}

The advantage of the present approach is that it allows to
estimate not only critical indices, but also the (non-universal)
thermodynamic quantities in a wide thermal range. Here we
calculate the dependence of the order parameter $\bar{x}$  on
temperature for the 3D Ising model. The model is defined as a
discrete cubic lattice with $U(x)=\delta(x-1)+\delta(x+1)$ and the
nearest-neighbors interaction $\Omega_k=3-\cos \pi k_x-\cos \pi
k_y-\cos \pi k_z$. The initial Brillouin zone is a cube
$-1<k_{x,y,z}<1$; at the first step the value $\bar{\Omega}=3$ is
used and all the modes with $|k|>1/s$ are integrated. The
procedure for further steps is exactly as described above. To
simplify the calculation, we put
$\bar{\Omega}^{(n)}=\bar{\Omega}^{(\infty)}$ at these steps; this
simplification almost does not change the result obtained. To
determine the value of $\bar{x}$ we use the condition
\begin{equation}\label{c_eta}
  \frac{\partial}{\partial x} U_n(\bar{x}_n)=0, n \to \infty
\end{equation}
(compare with the derivation of formula (\ref{MF})). Numerically,
it is enough to calculate $U_n$ up to $n \approx 8$.  The result
for $\bar{x}^2(T)$ is presented in Fig.2. The mean-field and
experimental (Monte-Carlo) data are sketched in the same figure
for comparison. Far from the transition $U^{(1)}$ is almost
parabolic, therefore the theory passes into MFA. The critical
behaviour occurs at the transition point. Note that Ising model is
very non-linear, but the scheme still performs good.

\begin{figure}
\includegraphics{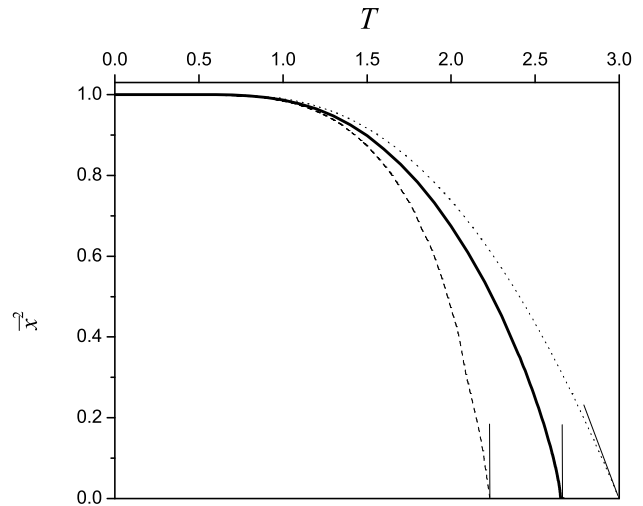}
\caption{\label{fig:fig2} The dependence of the square of the
order parameter on the temperature in the 3D Ising model. Solid
curve: the present approach with $s=2.5$; the dot curve:
mean-field calculation; the dash curve: numerical results. Thin
ticks show the tangents to $\bar{x}^2(T)$ in the transition
point.}
\end{figure}

We conclude that the presented low-order scheme describes
correctly all the physics of the second-order phase transitions in
3D systems with a short-range interaction and a scalar order
parameter. It gives a unified description both of the fluctuation
region and of the system far from the transition point. The method
does not pretend to be very accurate; its advantage consists in
simplicity. In principle, one can improve the accuracy by a
higher-order procedure for the integration in RG transformation.
An applicability of the method for another universality classes is
the subject of further studies.

The work was supported by RFFI foundation (grant 00-02-16253)  and
by the "Russian Scientific Schools" program (grant 96-1596476).

\end{document}